# Ion beam generated surface ripples: new insight in the underlying mechanism

*Tanuj Kumar,[1] Ashish Kumar[2], D. C. Agarwal,[1] N. P. Lalla,[3] and D. Kanjilal[1]*

[1]*Inter-University Accelerator Centre, Aruna Asaf Ali Marg, New Delhi-110067, India*

[2]*Indian Institute of Technology, New Delhi-110067, India*

[3]*UGC-DAE Consortium for Scientific Research, University Campus, Khandwa Road, Indore-452017, India*

A new hydrodynamic mechanism is proposed for the ion beam induced surface patterning on solid surfaces. Unlike the standard mechanisms based on the ion beam impact generated erosion and mass redistribution at the free surface (proposed by Bradley-Harper (BH) and its extended theories), the new mechanism proposes that the ion beam induced saltation and creep processes, coupled with incompressible solid flow in amorphous layer, leads to the formation of ripple patterns at the amorphous/crystalline (a/c) interface and hence at the free surface. Ion beam stimulated solid flow inside the amorphous layer controls the wavelength, where as the amount of material transported and re-deposited at a/c interface control the amplitude of ripples. The new approach is verified by designed experiments and supported by the discrete simulation method.

PACS: 81.16.Rf 68.35.Ct 68.49.Sf 79.20.Rf

Fabrication of self organized nano-structures over solid surfaces using energetic ion beam irradiation has received a remarkable attention in last couple of decades. It is an elegant and cost-effective single step approach over lithographic methods for device fabrication. In general a uniform ion irradiation of solid surfaces for intermediate energies ($10^2$–$10^4$ eV) causes a self-organized topographic pattern of ripples, holes or dots [1]. In 1988s, the first analytical approach to study the surface patterning was given by Bradley and Harper (BH) [2] on the basis of two competing processes: the destabilizing effect of curvature dependent roughening, and the stabilizing effect of surface diffusion. Further theoretical refinements of BH's model have been proposed to underline the secondary effect of local curvature dependent sputtering, ion beam induced smoothing and hydro-dynamical contribution [3-4]. BH's linear and its extended models explain many experimental observations but suffered many limitations also [5-7]. Investigations by Madi *et al*. [7] and Norris *et al*. [8] showed that the ion impact induced mass redistribution is the prominent cause of surface patterning and smoothening for high and low angles, respectively. Castro *et al.* [9-10] proposed the generalized framework of hydrodynamic approach, which considers ion impact induced stress causing a solid flow inside the amorphous layer. They pointed out that the surface evolution with ion beam is an intrinsic property of the dynamics of the amorphous surface layer [11]. All above experimental findings and their theoretical justification raise questions on lack of a single physical mechanism on origin and evolution of ripples on solid surface.

In this letter, we introduce a new approach for explaining all ambiguity related to origin of ripple formation. We argue that amorphous-crystalline interface (a/c) plays a crucial role in the evolution of ripples. It is proposed that the ion beam induced incompressible solid flow in amorphous layer starts saltation and creep of Si atoms at a/c interface which is responsible for ripple formation on the free surface rather than earlier mentioned models of curvature dependent erosion and mass redistribution at free surface. Experimental investigations together with theoretical simulations have been presented to show validity of our approach.

In order to study the role of a/c interface in surface patterning of Si (100) surface during irradiation, we performed a series of experiments by preparing two sets of samples with different depth location of a/c interface. The variation in depth location of a/c interface is achieved by irradiating the Si surface using 50 keV Ar$^+$ ion at a fluence of $5\times10^{16}$ ions/cm$^2$ (for full amorphization) at different angles of incidence viz. 60º (sample set-A) and 0º (sample set-B) with respect to surface normal. The depth location of a/c interface would be higher in set-B samples as compared to set-A samples due to higher projected ion range for $0^o$ as compared to $60^o$ ion beam irradiation. Fig. 1 (a) and (b) shows the schematic view for ion beam stimulated damage range for off-normal incidence of ion beam at $60^o$ (named as set-A) and normal incidence (named as set-B), respectively. Subsequently to grow ripples in second stage of irradiation, both sets of samples were irradiated at an angle of $60^o$ w. r. t. surface normal using 50keV Ar+ ion beam as shown in Fig. 1 (c & d). For the set-A samples ion beam stimulated damage effect will reach at a/c interface in $2^{nd}$ stage irradiation while it remains inside the amorphous layer for set B samples due to deeper depth location of a/c interface. To have detailed experimental data, a number of samples were prepared by varying fluence from $3\times10^{17}$ ions/cm$^2$ to $9\times10^{17}$ ions/cm$^2$ for each set. During the irradiation the base pressure of chamber was maintained at ~$10^{-7}$ mbar. The ion beam current density was kept constant at 15μA/cm$^2$. The beam was scanned uniformly over an area of 10 mm×10 mm by electromagnetic beam scanner. After irradiation, the samples were analyzed by Nano Scope IIIa atomic force microscope (AFM) under ambient conditions in tapping mode. Cross-sectional transmission electron microscopy (XTEM) was carried using a Tecnai-G2-20 TEM facility operating at 200 kV. The cross-sectional specimens for TEM study were prepared by using Ar ion-beam milling at 4 kV/20 μA and at an angle of $4^o$ with respect to the sample surface.

**Results**

AFM characterization was carried out on all samples after each irradiation step. After first irradiation, the average RMS roughness for both sets of the samples was nearly similar (0.5±0.1 nm and 0.6±0.1 nm, for set A and. In $2^{nd}$ stage, all samples were irradiated by a stable 50keV Ar$^+$ at same angle of incidence ($60^o$) for all fluences. Fig. 2(a-d) and (e-h) shows the AFM images for set-A and set-B samples after $2^{nd}$ stage irradiation at the fluences of $3\times10^{17}$ ions/cm$^2$, $5\times10^{17}$ ions/cm$^2$, $7\times10^{17}$ ions/cm$^2$ and $9\times10^{17}$ ions/cm$^2$ respectively. It was found that for set-A, the wavelength and amplitude were increasing with increase in irradiation fluence (as shown in Fig. 3(a,b)). For set-B samples, the average wavelengths of ripples were nearly same as that of set-A samples at corresponding fluences. However, the observed average amplitudes of ripples are about one order less in magnitude for set-B as compared to those for set-A. Since the only difference between two sets of samples was the depth location of a/c interfaces. If the evolution ripples were based on curvature dependent sputtering and surface diffusion, we should have got ripples of identical dimensions for corresponding equal fluence in both sets of samples. Despite similar initial surface morphology of both sets of samples after 1st stage of irradiation, the observation of similar wavelength and lower amplitude of ripples in set-B samples as compared to set-A samples cast doubt on the validity of Bradley-Harper and its extended theories. It can be emphasized here that we repeated complete set of experiment with two different ion beams and at different energies (Ar at 50keV and Kr at 60 keV). And in all cases the observed trend was similar. To the best of author's knowledge, there is no existing model which could physically explain this anomaly. The prominent role of the a/c interface in formation of ripples is established in this work.

Fig. 4(a), (b) and (c) show XTEM images for set-A samples corresponding to irradiation fluences of $5\times10^{16}$ ions/cm$^2$ (after first irradiation), $7\times10^{17}$ ions/cm$^2$ and $9\times10^{17}$ ions/cm$^2$, respectively. Similarly, Fig. 4(d) and (e) images are for set-B samples irradiated at fluences of $5\times10^{16}$ ions/cm$^2$ (after first irradiation), $7\times10^{17}$ ions/cm$^2$ , respectively. For the set-A samples (Fig. 4(a)), it was observed that top amorphous layer has a uniform thickness of about 74 nm which after irradiation at $7\times10^{17}$ ions/cm$^2$ results in ripple formation. From the XTEM images and using grid line method [12], it was found that during the rippling processes the overall cross-sectional area of amorphous layer remains constant which validates the condition of incompressible solid mass flow inside the a-Si layer [9-10]. For the set-B samples the initial a-Si layer thickness was found to be 170 nm as shown in Fig. 4(d). Interestingly, the thickness of a-Si was found to be decreased to 77 nm for the subsequent irradiated sample for the fluence of $7\times10^{17}$ ions/cm$^2$ (Fig. 4(e)). Observed ripple dimensions for all samples measured from XTEM were consistent with AFM data. Selected area diffraction (SAED) pattern taken on both sides of a/c-interface confirmed the amorphized and bulk crystalline regions as shown in Fig. 4(f).

To physically understand the underlying mechanism we considered a radical assumption that the formation of ripples is initiated at a/c interface due to erosion and re-deposition of Si atoms under the effect of solid flow. Due to incompressible nature of this solid mass flow inside amorphous layer, structures formed at the a/c interface reciprocate at top surface. Similar process of saltation in macroscopic ripple formation on sand (ripples caused by air flow on sand dunes etc.) has been well observed and studied [13-14]. Here, in our proposed model, we assume that the two elementary processes of saltation and creep of Si atoms are taking place at the a/c interface due to solid flow inside damaged layer, which controls the process of ripple formation.

In case of set-A samples the saltation and creep of Si atoms at the a/c interface starts instantaneously with second stage irradiation as the ion range is equal to depth of a/c interface. However, for set-B samples, second stage irradiation results in surface erosion before the ion beam effect reach at a/c interface. So, the process of saltation and creep (ripple formation) lag behind in set-B samples as compared to set-A samples. This fact was confirmed by the formation of ripples with appreciable average amplitude (23nm) and wavelength (780nm) observed at still higher fluence of $1.5\times10^{18}$ ions/cm$^2$. Therefore, in simulation we assumed that the base amount ($q_o$) of transported Si atoms would be small for set-B samples as compared to set-A samples for the fluence regime used.

**Saltation**: The process of saltation can be described by following equations [13]:

$$h'(x,y) = h(x,y) - q_s(x,y) \tag{1}$$

$$h'(x+l_{sx}, y+l_{sy}) = h(x+l_{sx}, y+l_{sy}) + q_s(x,y) \tag{2}$$

$$\vec{l}_s(x,y) = l_{sx}(x,y)\vec{i} + l_{sy}(x,y)\vec{j} \tag{3}$$

$$q_s(x,y) = q_0 + c\tanh(g(x,y)) \tag{4}$$

Where $h(x,y)$ and $h'(x,y)$ are the heights of silicon surface at a point (x,y) on a/c interface before and after saltation movement, respectively; $\vec{l}_s(x,y)$ represents the horizontal saltation displacement vector, $q_s(x,y)$ stands for the transferred height of Si moving one step, $q_0$ is a control parameter corresponds to base amount of silicon, $g(x,y)$ is a function of the gradient of a/c surface at point (x,y), defined as:

$$g(x,y) = sign\left(\frac{\partial h}{\partial x}\right)\sqrt{\left(\frac{\partial h}{\partial x}\right)^2 + \left(\frac{\partial h}{\partial y}\right)^2} \tag{5}$$

Saltation of Si atoms at a/c interface is caused by solid flow inside the amorphous layer. The silicon atoms are picked up by the solid flow and deposited some distance later that is proportional to the solid flow velocity. Castro *et al.* [9-10] and Kumar *et al.* [12] have also discussed role of solid flow for surface rippling. Kumar *et al.* had experimentally shown that solid flow velocity depends on the local curvature of a/c interface. From our XTEM results also it is found that during the rippling processes the overall cross-sectional area of amorphous layer remains constant which validates the condition of incompressible solid mass flow inside the a-Si layer. With the assumption of incompressible solid flow inside the amorphous layer, we define the solid flow velocity inside the amorphous layer ($V_s^{x,y}$) and saltation length ($l_{s(x,y)}(x,y)$) as:

$$V_s^{x,y} = V_o[1 - \tanh(\frac{\partial h}{\partial x})] \tag{6}$$

$$l_{s(x,y)}(x,y) = (l_0)V_s^{x,y} \tag{7}$$

where $V_o$, $l_0$ are control parameters, and $\frac{\partial h}{\partial x}$ and $\frac{\partial h}{\partial y}$ are the slopes in x and y directions.

**Creep/smoothing:** Creep is the smoothening of ripples which settles and flatten the a/c interface, and can be described by:

$$h_{n+1}(x,y) = h_{n'}(x,y) + D\left[\frac{1}{6}\sum_{NN}h_{n'}(x,y) + \frac{1}{12}\sum_{NNN}h_{n'}(x,y) - h_{n'}(x,y)\right] \tag{8}$$

Where, $\sum_{NN}h_{n'}(x,y)$ and $\sum_{NNN}h_{n'}(x,y)$ represents the summation of the heights of the adjacently horizontal, vertical and diagonal sites; and D is a relaxation factor [14].

Simulations were carried out using the above formulism for different run steps at two values of base amount control parameter $q_o$. Since the initial velocity ($V_o$) and length ($l_o$) control parameters depend on the ion beam irradiation parameters, so they are taken same for two cases. Variation of simulation run steps denotes variation of irradiation fluence values in actual experiment. Fig. 5 (a and b) shows the simulation results for the parameters of $V_o$=20, $l_o$=0.5, a=0.1, c=0.1, $q_o$ =0.1 for 60 and 200 run steps. While, Fig. 5 (c and d) shows the simulation results for the parameters of $V_o$=20, $l_o$=0.5, a=0.1, c=0.1, $q_o$ =0.5 for 60 and 200 steps. Fig. 6 (a) and (b) show variation of wavelength and amplitude with run steps at two values of $q_o$. From the simulation results we find that the average wavelength of ripples is nearly same for each corresponding run step. However the observed growth rate in amplitude is much higher for $q_o$=0.5 as compared to $q_o$=0.1. The theoretically simulated results (Fig. 6) are in well agreement with experimentally observed (Fig. 3) data. This result shows that the wavelength of ripples is a function of solid flow velocity while the amplitude is a function of base amount to be transported or re-deposited at a/c interface. This saltation length controls the ripple wavelength as also observed in macroscopic world [13-14]. As already discussed, our AFM and XTEM results could not be explained by existing models of BH and its extended theories, where they consider it only surface effect. Role of a/c interface have not been considered in formation of ripples on solid surfaces by earlier groups [2, 8-9]. By considering ripple formation as a/c interface dependent process, all phenomena like ripple coarsening, propagation etc. can be correlated.

In conclusion, by designed experiments and theoretical modeling, a new approach for explaining origin of ripple formation on solid surface has been proposed. Formation of ripples at top surface is a consequence of saltation process at the a/c interface induced by incompressible solid flow inside the amorphous layer. The control parameter for ripple wavelength is solid flow velocity, while that for the amplitude is amount of silicon to be transported at the interface. The contribution from the ion beam induced curvature dependent sputtering and mass redistribution on the free surface proposed by BH theory are essentially insignificant in surface patterning.

**Acknowledgement**

One of the authors (Tanuj Kumar) is thankful to Council of Scientific and Industrial Research (CSIR), India for financial support through senior research fellowship. The help received from S. A. Khan and U. K. Rao during the experiment is gratefully acknowledged here.

**Figure Captions**

FIG. 1 Schematic view of 50 keV $Ar^+$ ion beam irradiation for $1^{st}$ stage (to prepare two deferent depth locations of a/c interface) at an angle of (a) 60° and (b) 0°, with respect to surface normal. $2^{nd}$ stage irradiation (for fabrication of ripples) at an angle of 60° named as (c) set-A and (d) set-B.

FIG. 2 AFM images for the 50 keV $Ar^+$ irradiated set-A and set-B samples at an angle of 60° with respect to surface normal at the fluences of $3\times10^{17}$ ions/cm$^2$ (a,e), $5\times10^{17}$ ions/cm$^2$ (b,f), $7\times10^{17}$ ions/cm$^2$ (c,g), $9\times10^{17}$ ions/cm$^2$ (d,h), respectively. The arrows in the figures indicate the projection of ion beam direction on the surface.

FIG. 3 Variation of wavelength (a) and amplitude (b) of ripples for set-A and set-B samples with ion beam fluence.

FIG. 4 X-TEM images of 50 keV $Ar^+$ irradiated set-A samples at the fluences of (a) $5\times10^{16}$ ions/cm$^2$, (b) $7\times10^{17}$ ions/cm$^2$, (c) $9\times10^{17}$ ions/cm$^2$, and set-B samples (d) $5\times10^{16}$ ions/cm$^2$ (for normal incidence), (e) $7\times10^{17}$ ions/cm$^2$. SAD pattern for the amorphized and bulk crystalline regimes.

FIG. 5 Simulation results for parameters $V_s$=20, $l_o$=0.5, a=0.1, c=0.1, $q_o$=0.1 for run steps (a) 60 (b) 200, and $V_s$=20, $l_o$=0.5, a=0.1, c=0.1, $q_o$=0.5 for run step (c) 60 and (d) 200.

FIG. 6 Simulated variation of wavelength (a), and amplitude (b), of ripples for two base amount of $q_o$ = 0.1 and 0.5.

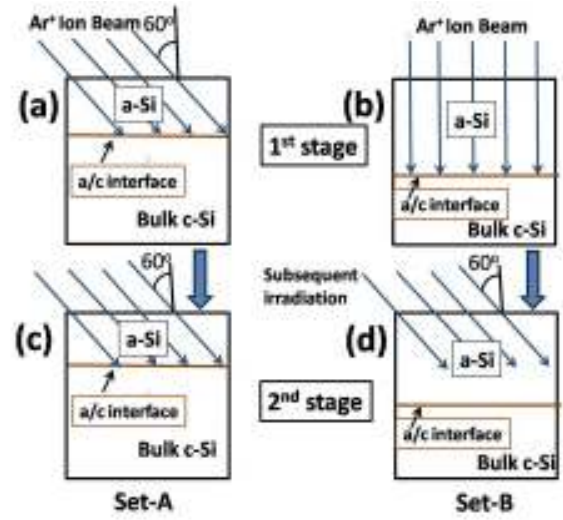

FIG. 1

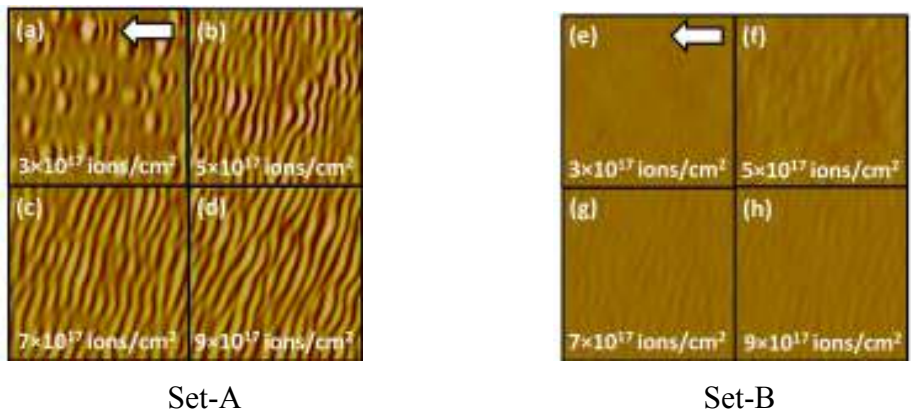

Set-A                                      Set-B

FIG. 2

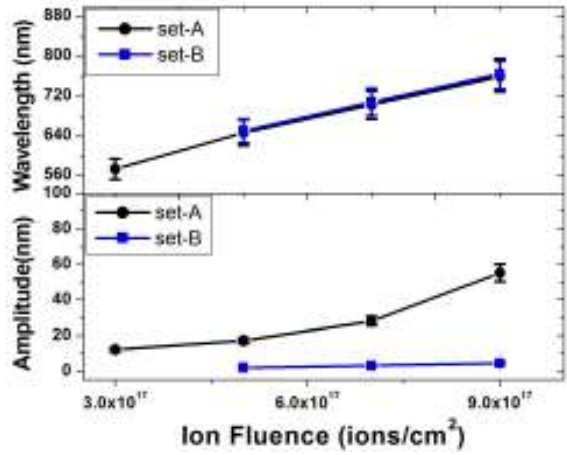

FIG. 3

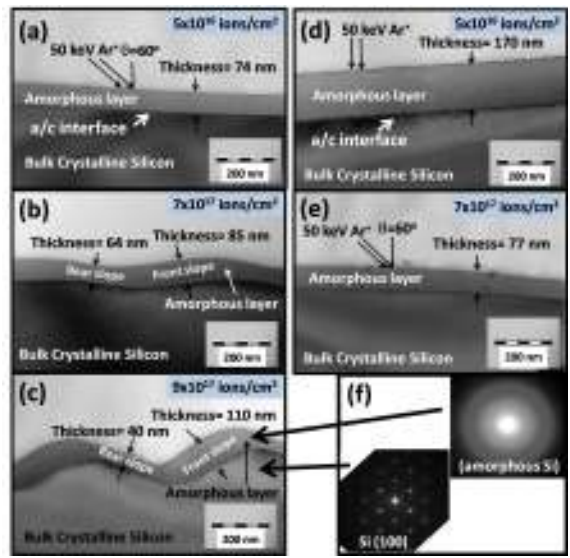

Fig. 4

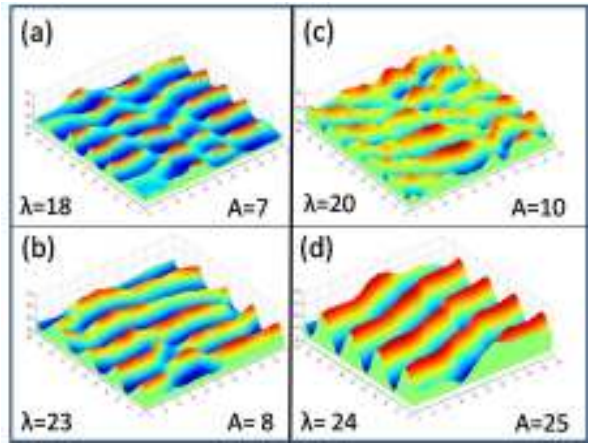

Fig. 5

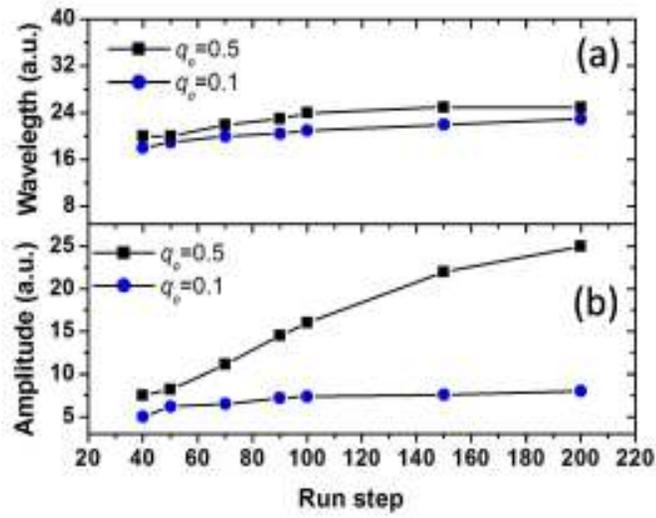

Fig. 6